\documentclass[english,aps,prl,superscriptaddress,floatfix, notitlepage,reprint]{revtex4-1}
\usepackage[T1]{fontenc}
\usepackage[latin9]{inputenc}
\usepackage{amsmath}
\usepackage{amsthm}
\usepackage{amssymb}

\makeatletter

\usepackage{braket}
\usepackage{txfonts} 
\usepackage{graphicx}
\usepackage[usenames,dvipsnames]{xcolor}
\usepackage[colorlinks=true,citecolor=Blue,linkcolor=RubineRed,urlcolor=Blue]{hyperref}
  \hypersetup{
           breaklinks=true,   
           colorlinks=true,   
           pdfusetitle=true,  
        }
\date{\today}

\makeatother
\usepackage{babel}

\usepackage{float}

\def\la{\lambda}

\def\N{{\rm N}}

\def\la{\langle}
\def\ra{\rangle}
\def\om{\omega}

\def\tr{{\rm Tr}}

\newcommand{\beq}{\begin{equation}}
\newcommand{\eeq}{\end{equation}}
\newcommand{\beqa}{\begin{eqnarray}}
\newcommand{\eeqa}{\end{eqnarray}}

\begin{document}
\title{Tailoring Dynamical Fermionization: Delta kick cooling of a Tonks-Girardeau gas}
\author{L\'eonce Dupays \href{https://orcid.org/0000-0002-3450-1861}{\includegraphics[scale=0.05]{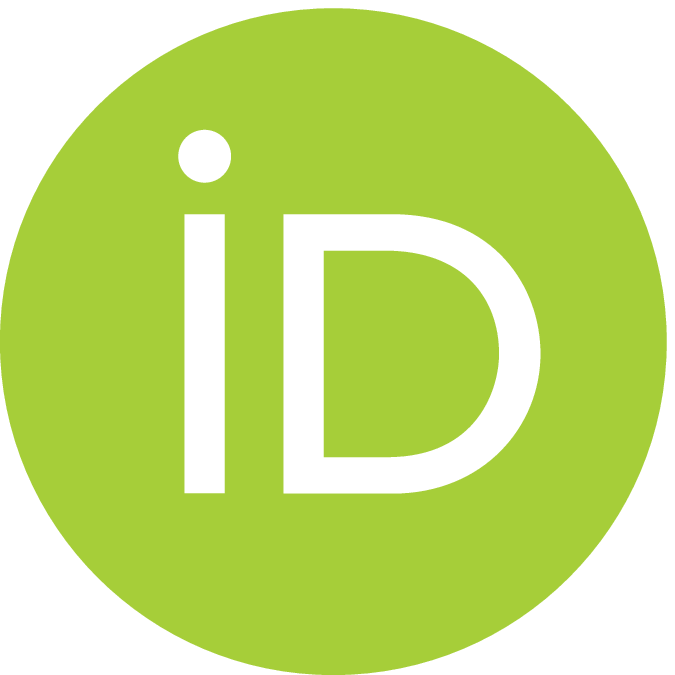}}}
\email{leonce.dupays@uni.lu}
\affiliation{Department of Physics and Materials Science, University of Luxembourg, L-1511 Luxembourg, G. D. Luxembourg}
\author{Jing Yang \href{https://orcid.org/0000-0002-3588-0832}{\includegraphics[scale=0.05]{orcidid.eps}}}
\email{jing.yang@uni.lu}
\affiliation{Department of Physics and Materials Science, University of Luxembourg, L-1511 Luxembourg, G. D. Luxembourg}
\author{Adolfo del Campo  \href{https://orcid.org/0000-0002-3450-1861}{\includegraphics[scale=0.05]{orcidid.eps}}}
\email{adolfo.delcampo@uni.lu}
\affiliation{Department of Physics and Materials Science, University of Luxembourg, L-1511 Luxembourg, G. D. Luxembourg}
\affiliation{Donostia International Physics Center, E-20018 San Sebasti\'an, Spain}

\begin{abstract}
In one spatial dimension, quantum exchange statistics and interactions are inextricably intertwined. As a manifestation, the expansion dynamics of a Tonks-Girardeau gas is characterized by dynamical fermionization (DF), whereby the momentum distribution approaches that of a spin-polarized Fermi gas. Using a phase-space analysis and the unitary evolution of the one-body reduced density matrix, we show that DF 
can be tailored and reversed, 
using a generalization of delta kick cooling (DKC) to interacting systems, establishing a simple protocol to rescale the initial momentum distribution. The protocol applies to both expansions and compressions and can be used for microscopy of quantum correlations.
\end{abstract}
\maketitle

In one spatial dimension, the interchange of particle coordinates inevitably leads to their interactions. In the process, the quantum state accumulates a phase shift that is the sum of two contributions, one stemming from quantum statistics, and the other from scattering. 
This fact makes it possible to relate the physics of some strongly correlated quantum gases to that of noninteracting systems with different quantum statistics. 
This is the basis of the Bose-Fermi duality introduced by Girardeau in 1960 to describe a gas of one-dimensional hard-core bosons, which is now known as the Tonks-Girardeau (TG) gas \cite{Girardeau60, Cazalilla11}. The latter can be described in terms of a spin-polarized one-dimensional Fermi gas with no interactions. Pauli exclusion principle in the Fermi gas makes the wavefunction vanish at contact, a feature shared by the TG gas due to the hard-core interactions. 
The wavefunctions of the two systems are identical for a given particle ordering and differ only in their symmetrization. The bosonic TG wavefunction $\Psi_{\rm TG}$ can be obtained from that of the Fermi gas $\Psi_{\rm F}$ by explicit symmetrization according to the Bose-Fermi mapping $\Psi_{\rm TG}=\prod_{j<k}{\rm sgn}(x_k-x_j)\Psi_{\rm F}$. 
Similar relations exist in systems governed by strongly-attractive $p$-wave interactions \cite{Granger04,Girardeau04,GirMin06}. 
One-dimensional dualities can be extended to general exchange statistics, in and out of equilibrium \cite{Girardeau06,delcampo08}, paving the way to the experimental realization of hard-core anyons \cite{Keilmann11}. They can be further used to describe systems with finite-interaction strength \cite{Buljan08,Batchelor08,Xiwen2013}, mixtures \cite{GirMin07}, and spinor systems \cite{Deuretzbacher08}, among other examples \cite{Cazalilla11,GuanRMP13}.

The TG gas can be considered as the strong-coupling limit of the Lieb-Liniger (LL) gas, which describes one-dimensional bosons subject to contact interactions of finite-strength $c$ \cite{LL63, L63, Olshanii98}.  
 This model is integrable and solvable by Bethe ansatz. 
 The relevance of the LL gas to ultracold atom physics was established by Olshanii, who showed that ultracold atoms in tight waveguides are described by the LL model with tunable coupling constant $c$ \cite{Olshanii98}.
The strongly interacting limit $c\rightarrow+\infty$ leads to the TG regime,  realized experimentally by making use of an optical lattice \cite{Kinoshita04, Kinoshita05, Cazalilla04}.
The connection between the continuum and lattice version of hard-core bosons is well-understood \cite{Cazalilla04}, and dynamical correlations are in one-to-one correspondence at low densities. 

Local correlations such as the density profile are indistinguishable between dual systems. By contrast, correlations depending on the off-diagonal elements of the density matrix exhibit clear signatures of quantum statistics \cite{PenroseOnsager56, Lenard64}. A prominent example is the momentum distribution. While that of a 1D Fermi gas in the ground state exhibits a characteristic flat profile, that of the TG gas is sharply peaked at $k=0$ and has power-law tails decaying as $1/k^4$.  
 Out of equilibrium, it was predicted that a TG gas under free expansion in 1D exhibits dynamical fermionization (DF), with the asymptotic momentum distribution of the TG gas matching that of free fermions \cite{RigolMuramatsu05, MinguzziGangardt05}. 
This phenomenon has been recently observed in the laboratory for the first time \cite{Wilson1461}.
DF also governs the asymptotic behavior of an expanding LL gas, which enters the TG regime \cite{Jukic08, Buljan09}. While it is conveniently described using scale-invariance, which makes the density profile at different times self-similar, it does not rely on it, and occurs whether the initial confinement is harmonic or not \cite{delcampo06,delcampo08,Campbell15}. Generalizations of this phenomenon have been reported for a fermionic analog of the TG gas \cite{GirMin06}, hardcore anyons \cite{delcampo08}, and spinor quantum gases \cite{Alam21}.
DF is generally justified as a result of free expansion along the axial direction: as the particle density decreases, the asymptotic momentum distribution is that of the rapidities, which are the conserved quantities in a many-body integrable quantum system \cite{Sutherland98,RigolMuramatsu05,MinguzziGangardt05,delcampo06,delcampo08,Bolech12,Campbell15, Mei16}.

In this Letter, we analyze DF in phase space in arbitrary scale-invariant processes, showing that its appearance is not restricted to expansions but can occur as well in an implosion protocol leading to a density increase. While the momentum distribution and density profile of dual systems under DF become equal, the one-body reduced density matrix is shown to evolve unitarily, making the distinguishability of the corresponding quantum states independent of time. As a result, DF can be reversed, making use of a generalization of delta kick cooling (DKC) to interacting systems, pulsing an external potential.
This allows to engineer protocols that rescale the momentum distribution for microscopy of quantum correlations.

{\it TG gas in a time-dependent trap.---}
Consider a TG gas in a harmonic trap, dual to an ideal Fermi gas in the same confinement \cite{Girardeau01}. In the ground state, the TG wavefunction is the absolute value of the fermionic one, which is given by a Slater determinant, e.g., $\Psi_{\rm F}(x_1,\dots,x_\N)=\frac{1}{\sqrt{\N!}}\det_{n=0,k=1}^{\N-1,\N}[\phi_n(x_k)]$ in terms of the single-particle harmonic oscillator eigenstates. Both systems are scale invariant with dimensionality $D=1$, and their time-dependent coherent states take the form \cite{Sutherland98,MinguzziGangardt05,Gritsev10,delcampo11}
\beqa
\label{psit}
\Psi\left(t\right)&=&
\frac{1}{b^{\frac{\N}{2}}}\exp\left[i\frac{m\dot{b}}{2\hbar b}\sum_{i=1}^\N x_i\,^2-i\int_{0}^t\frac{E(0)}{\hbar b(t')^2}dt'\right]\nonumber\\
& & \times \Psi\left(\frac{x_1}{b},\dots,\frac{x_\N}{b},t=0\!\right)\,,
\eeqa
where the scaling factor $b(t)>0$ is the solution of the Ermakov equation $\ddot{b}+\om(t)^2b=\om_0^2/b^{3}$ with the initial conditions $b(0)=1$, $\dot{b}(0)=0$. 
Note that this scaling law is not restricted to the ground-state but it is shared by any many-body eigenstate $\Psi(0)$ with energy eigenvalue $E(0)$.
Quantities derived from $|\Psi(t)|^2$ are shared by dual systems related by the Bose-Fermi mapping, given that $\prod_{j<k}[{\rm sgn}(x_k-x_j)]^2=1$. By contrast, those dependent on the coherence in real space generally differ.
We focus on the one-body reduced density matrix (OBRDM), that contains all the information required to analyze one-body observables. It is defined as
$\rho_1(x,x',t)=\N\int dx_2\cdots dx_\N\Psi\left(x, x_2\cdots x_\N,t\right)\Psi^*\left(x', x_2\cdots x_\N,t\right)$. From it, one can determine the density profile
$\rho(x,t)=\rho_1(x,x,t)$, as well as the momentum distribution, making use of the Fourier transform
 $n(p,t) = \frac{1}{2\pi\hbar}\int dxdx' e^{-ip(x-x')/\hbar} \rho_1(x,x',t)$.
Using (\ref{psit}), the OBRDM evolves according to
\beqa
\rho_1(x,x',t)
=\frac{1}{b}\exp\left[i\frac{m\dot{b}}{2\hbar b}(x^2-x'\,^2)\right]\rho_1\left(\frac{x}{b},\frac{x'}{b},t=0\right)\label{eq:rhot}.
\eeqa
Similar relations hold for the time-evolution of higher-order reduced density matrices. 
In the limit of adiabatic driving $\dot{b}/b\to 0$, the OBRDM is rescaled as $\rho_1(x,x',t)= \rho_1(x/b,x'/b,0)/b$.
The use of controlled expansions involving time-dependent traps and engineered by shortcuts to adiabaticity has been proposed for implementing such scaling without the requirement of slow driving, but generally involve time-dependent traps \cite{delcampo11,delcampo12}. Such protocols realize in essence a dynamical microscope zooming in on correlations in the OBRDM. \\

{\it Shared unitary evolution of the OBRDMs and its consequences.---}Interestingly, Eq.~\eqref{eq:rhot} indicates that the evolution of OBRDMs for both the TG gas and the spin-polarized ideal Fermi gas is unitary. More precisely, we introduce a generic label $A=\{{\rm TG,F}\}$ for any of the dual systems and define the corresponding quantum state $\sigma^A=\frac{1}{\N}\int dxdx'\rho_1^{A}(x,x')|x\ra\la x'|$ such that ${\rm Tr}(\sigma^A)=1$. Then Eq.~\eqref{eq:rhot} implies $\sigma^A(t)=U(t)\sigma^A(0)U^\dag(t)$, where
\beqa
U(t)=\exp\left[i\frac{m\dot{b}}{2\hbar b}x^2\right]\exp\left[-i\frac{{\rm ln}b}{2\hbar}(xp+px)\right], 
\eeqa
and the rightmost term is the dilatation operator implementing a scaling transformation in real space by a factor $b$. Note that $U(t)$ is the same for both $A=\{{\rm TG,F}\}$. In fact, under scale invariant dynamics, 
the evolution of the quantum state associated with any $s$-body reduced density matrix is also unitary~\cite{SM}.

The identical unitary evolution of the OBRDMs has several consequences: (i) The spectral decomposition of the OBRDM is of the form $\rho_1(x,x',t)=\sum_\mu\lambda_\mu(0)\phi_\mu(x,t)\phi_\mu(x',t)$, where  
 the eigenvalues $\lambda_\mu(0)$ of the OBRDM, which correspond to the occupation numbers of the natural orbitals $\phi_\mu(x,t)$, are constant in time. 
It follows that the dynamics is isentropic, i.e., it preserves the von Neumann entropy $S(\sigma^A)=-\tr[\sigma^A\log\sigma^A]$.
The situation in the continuum is thus in contrast with that reported for hard-core bosons in an optical lattice \cite{RigolMuramatsu05}. 
In addition, the natural orbitals fulfill the relation $\phi_\mu(x,t)=\exp\left[i\frac{m\dot{b}}{2\hbar b}x^2\right]\phi_\mu(x/b,0)/\sqrt{b}$. (ii) Consider the Ulhmann fidelity defined as  $\mathcal{F}(\sigma, \sigma')={\rm Tr}\left(\sqrt{\sqrt{\sigma}\sigma'\sqrt{\sigma}}\right)$ \cite{Uhlmann92,Nielsen00} as a distinguishablity measure between the quantum states $\sigma$ and $\sigma'$. 
Given that the Uhlmann fidelity is invariant under conjugation of its arguments by a common unitary, it follows that
 \begin{equation}
 \mathcal{F}[U(t)\sigma^{\rm TG}(0)U^\dag(t), U(t)\sigma{^{\rm F}}(0)U^\dag(t)]= \mathcal{F}[\sigma^{\rm TG}(0),\sigma^{\rm F}(0)] \label{eq:eq-fid}.
\end{equation}
As a result, even if the density profile, and the momentum distribution under DF, are shared by both dual systems, their quantum states remain equally distinguishable at all times. (iii) The unitary evolution in quantum mechanics can be represented by a kernel in the phase space, convoluted with the initial Wigner function of the system~\cite{GarciaCalderon80}. Under scale-invariant dynamics, the convolution simplifies to a linear canonical transformation on the initial Wigner function~\cite{SM}, which motivates the phase-space analysis of DF. 
(iv) Finally, since the unitary evolution is invertible, this paves the way to reverse DF, which we shall discuss in what follows.

{\it Phase-space analysis of DF.---}  
The Wigner function associated with the OBRDM can be represented as a function of the coordinate $x$ and the canonically-conjugated momentum $p$ \cite{Wigner32,Hillery84}, 
\beqa
W(x,p,t)= \frac{1}{\pi\hbar} \int_{-\infty}^\infty 
\rho_1(x-y,x+y,t)
 e^{2i py/\hbar} d y\,. \label{eq:wigner_function}
\eeqa
The marginals of $W(x,p,t)$ correspond to the density profile $\rho(x,t)=\rho_1(x,x,t)=\int W(x,p,t)dp$ and the momentum distribution $n(p,t)=\rho_1(p,p,t)= \int W(x,p,t)dx$. From the dynamics (\ref{eq:rhot}) of the OBRDM, following an arbitrary modulation of the trapping frequency $\om(t)$, 
the exact time-evolution of the Wigner function reads \cite{Shanahan18,SM} 
\beqa
\label{WtSI}
 W(x,p,t)&=&W\left(\frac{x}{b},bp-m\dot{b}x,t=0\right), \label{eq:rescaling}
 \eeqa
 where we note that $W$ does not need to be positive, i.e., it can describe a non-classical state. 
Note that this evolution is common to all scale-invariant systems, e.g., such as the single-particle time-dependent harmonic oscillator.
The Wigner function is rescaled, stretching (compressing) the density profile along the $x$-axis, and compressing (stretching) the momentum distribution along the $p$-axis if $b(t)>1$ ($b(t)<1$). The supplementary term $-m\dot{b}x$ involves a shift in phase-space, which induces DF.

\begin{figure}[t]
\begin{centering}
\includegraphics[width=0.9\linewidth]{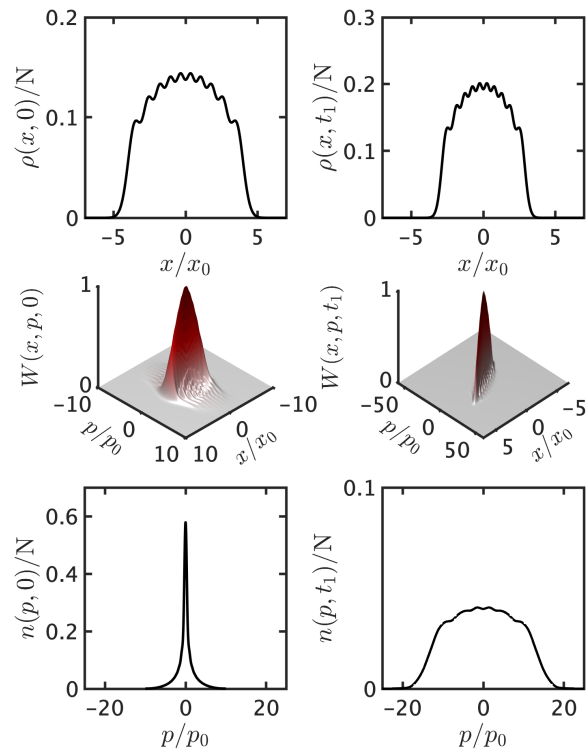}
\caption{DF of a TG gas in a sudden expansion in phase space.
Under scale-invariant dynamics, DF results from a canonical transformation that describes the evolution of the Wigner function. The density profile is scale-invariant at all times. In addition, when the rate of change of the scaling factor is large, the asymptotic momentum distribution can be related to the initial density profile.
The initial Wigner function is peaked along the axis $p=0$, with riddles on both sides that take negative values.
We chose $\omega_{0}=5\omega_{1}$. At time $t_{1}=\frac{3\pi}{4\omega_{1}}$ the Wigner function is rotated and dilated in the phase-space, so that the momentum distribution corresponds to the rescaled density profile. We chose $\N=10$.\label{Fig: Wigner_density_momentum}}
\end{centering}
\end{figure}

 Both for the TG and Fermi gas, the density profile exhibits explicitly the scale invariance, while the asymptotic momentum distribution can be related to the initial density profile \cite{SM}:
 
 \begin{eqnarray}
 \rho(x,t)=\frac{1}{b}\rho\left(\frac{x}{b},0\right),
 n(p,t)\approx\frac{1}{m\dot{b}}\rho\left(\frac{p}{m\dot{b}},0\right). \label{eq:DF_equation}
 \end{eqnarray}
 Note that the relation between $n(p,t)$ and $\rho(x,0)$ is different from what is known in time-of-flight imaging, connecting $\rho(x,t)$ to $n(p,0)$.
 The results in (\ref{eq:DF_equation}) are consistent with previous studies limited to sudden expansions \cite{Buljan09,Campbell15}. Note the first equation of  Eq.~(\ref{eq:DF_equation})  is exact while the second is approximate, which requires that the initial Wigner function decays over the characteristic spread $\Delta p$ so that during the dynamics $\Delta p\ll x_{0}m \dot{b}b$ with $x_0=\sqrt{\hbar/(m\omega_0)}$, which is equivalent to the condition $\dot{b}b\gg 2\omega_{0}$ taking $\Delta p \approx 2 p_{0}$, where $p_0=\hbar/x_0$.  For a rigorous asymptotic analysis, see~\cite{SM}. In the special case of a harmonic trap, the initial density profile of the TG gas (and its dual system, the spin-polarized Fermi gas) can be expressed in terms of the rescaled momentum distribution of the Fermi gas \cite{SM}, making the term ``DF'' natural in this setting  \cite{RigolMuramatsu05, MinguzziGangardt05}. 
For the sake of demonstration, we consider the expansion of a TG gas initially confined in a harmonic trap with frequency $\omega_{0}$ that is suddenly released in a wider trap with frequency $\omega_{1}$, illustrated in Fig. \ref{Fig: Wigner_density_momentum}. 
 This process leads to periodic time-dependence of the scaling factor $b(t)$ \cite{MinguzziGangardt05,dupays2021}; see \cite{SM} for other protocols. The width of the cloud is controlled by $b(t)$ and oscillates after the release of the TG gas into the wider trap. This behavior induces DF periodically, with the nonequilibrium momentum distribution evolving between that of an equilibrium TG gas and a Fermi gas, as predicted in \cite{MinguzziGangardt05} and recently observed experimentally \cite{Wilson1461}. 
As emphasized, the phase-space dynamics in Eq. (\ref{WtSI})  holds for all scale-invariant systems and is responsible for the asymptotic form of the momentum distribution can be related to the initial density profile, whenever the width of the atomic cloud varies swiftly. The fact that the asymptotic momentum distribution can be also exactly related to that of the spin-polarized Fermi gas is a specific feature of the harmonic confinement, as single-particle eigenstates $\phi_n(x)$ are in this case expressed in terms of Hermite polynomials, that are eigenstates of the Fourier transform. Note that despite the coincidence of the marginals of the Wigner functions of the TG gas and the Fermi gas at late times, the OBRDMs remain distinguishable as discussed above. 

{\it Tailoring and reversing DF with kicks.---} The momentum shift in (\ref{WtSI}) is responsible for DF. Classically, one may expect to cancel it by applying a conservative force for a short period of time $\tau_k$ inducing a momentum change $\delta p=-\tau_k\partial_xV(x)$, i.e., pulsing an external potential $V(x)$. This argument, limited to classical noninteracting systems, is the basis of delta-kick cooling (DKC) \cite{Chu86,Ammann97,Morinaga99}. In what follows, we make use of the extension of DKC for scale-invariant interacting systems. Under Eq. (\ref{eq:rhot}), excitations encoded in the phase factor proportional to $\dot{b}/b$, that are responsible for DF, can be explicitly canceled in an interacting system by applying a kick potential of appropriate strength. Canceling the phase allows to tailor the momentum distribution and reverse DF.  Given that the phase oscillation in Eq. (\ref{psit}) is quadratic in the coordinates it can be canceled by pulsing an external harmonic trap with a given frequency $\om_k$.
To this end, consider the 
Hamiltonian with a $\delta$-kick applied at $t_k$
\beqa
\label{eq:delta}
H_k(t)=H(t)+\delta(t-t_k)\frac{1}{2}m\om_k^2\sum_{i=1}^{\N} x_i\,^{2}.
\eeqa
The use of a delta function is justified when the duration of the pulse $\tau_k$ is short with respect to other time scales \cite{dupays2021}.
The corresponding time-evolution operator admits the factorization
$U_{ \delta}(t_{F},0)=e^{-i\tau_k\frac{m\om_k^2}{2\hbar}\sum_{i=1}^{\N} x_i\,^{2}}U(t_k,0)$,
where $U(t,t')$ is the propagator associated with $H(t)$, and $\tau_k$ is a small time scale during which the kick is applied. 
Considering the evolution from $t=0$ to time $t_k+\tau_k$, one can choose the pulse parameters $\tau_k$ and $\om_k$ 
such that
\beqa
\tau_{k}\omega^{2}_{k}=\frac{\dot{b}(t_{k})}{b(t_{k})}\,.
\label{DKCcond}
\eeqa
This requires pulsing a harmonic trap with $\om_k>0$ in an expansion with $\dot{b}(t_{k})>0$ and an inverted harmonic trap with purely imaginary frequency $i\omega_k$ in a compression with $\dot{b}(t_{k})<0$. 
 In either case, the application of the kick cancels DF, and brings back the OBRDM to the initial one up to a scaling of the coordinates with respect to $b$, 
\beqa
\rho_1(x,x',t_k+\tau_k)&=&\frac{1}{b(t_k)}\rho_1\left(\frac{x}{b(t_k)},\frac{x\,'}{b(t_k)},t=0\right).
\eeqa
Consequently, the momentum distribution after the kick is $ n(p,t)=b\, n(pb,0)$, and similarly the Wigner function reduces to $W(x,p,t)=W_0(x/b,bp,0)$ with $b=b(t_k)$.
In turn, DKC prepares the same state that would have been obtained under adiabatic dynamics, without the requirement of slow driving.  

\begin{figure}[t]
\begin{centering}
\includegraphics[width=1\linewidth]{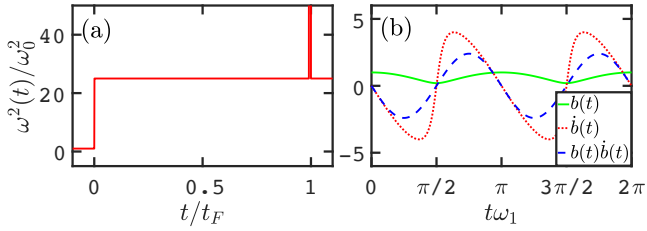}
\caption{ (a) The frequency modulation leading to the implosion protocol relies on a sudden compression from the initial frequency $\om_0$ to the final frequency $\omega_{1}/5=\omega_{0}=1$. DF is reversed by making use of pulsed attractive harmonic potential. For illustration, we chose the frequency of the kick $\omega^{2}_{k}=1000\omega^{2}_{0}$, $t_{k}=3\pi/4\omega_{1}$ and the final frequency determined by $b_{F}=\sqrt{\omega_{0}/\omega_{F}}=\sqrt{1/5}$. (b) Evolution of the scaling factor, its derivative, and the product $b\dot{b}$. 
\label{MainFig2}}
\end{centering}
\end{figure}

\begin{figure}
\begin{centering}
\includegraphics[width=0.9\linewidth]{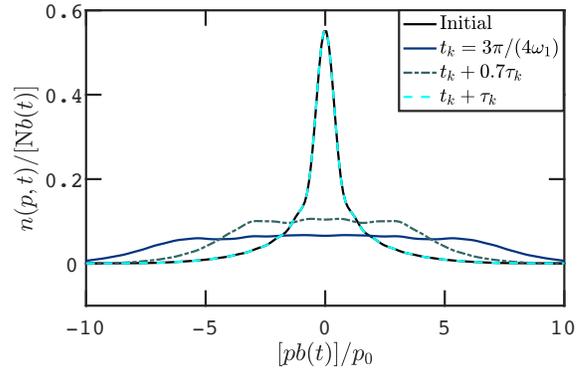}
\caption{DKC of an imploding TG gas. (a) Momentum distribution of a trapped TG gas at $t=0$, after the implosion at $t_{k}=\frac{3\pi}{4\omega_{1}}$, after the implosion followed by a  pulse of shorten duration $0.7 \tau_{k}$ (green dashed-dotted line), and with the DKC duration $\tau_{k}$ satisfying (\ref{pulsep}) (turquoise blue dashed line); $\N=7$. In dashed green the kick strength is chosen as $\tau_{k}\omega^{2}_{k}$ in order to cancel DF, with $\omega^{2}_{k}=1000\omega^{2}_{0}$. In choosing the strength of the kick differently one can modulate DF. The implosion protocol is corresponds to the sudden compression of the trap from $\omega_{0}$ to $\omega_{1}=5\omega_{0}$. \label{MainFig3}}
\end{centering}
\end{figure}

 {\it Imploding TG gas.---} Intuitively the DF occurs for an expansion as the particle density decreases. 
 This is the case considered so far in theoretical and experimental studies, in which it is possible to suppress DF by DKC as we show in \cite{SM}.  
 Yet, the phase-space dynamics (\ref{WtSI}) also yields DF in a compression process with $b(t_{F})<b(t=0)=1$ if the rate of change of the scaling factor is fast enough so that $\dot{b}b$ is large. For the sake of illustration, we consider a sudden compression protocol, where the trap of initial frequency $\omega_{0}$ is compressed to a frequency $\omega_{1}>\omega_{0}$, leading to the periodic scaling factor $b(t)$ displayed in Fig. \ref{MainFig2}. Large values of $\dot{b}b$ induce a high-frequency phase modulation in the coordinate representation. 
%
%
For the prescribed protocol, applying at the time $t_k$ a kick of duration $\tau_k$, the required pulse parameters to reverse DF are set by
\beqa
 \tau_{k}\omega^{2}_{k}=\frac{\omega_{1}\left(\omega^{2}_{0}-\omega^{2}_{1}\right)\sin(\omega_{1}t_{k})\cos(\omega_{1}t_{k})}{\left(\omega^{2}_{0}-\omega^{2}_{1}\right)\sin^{2}(\omega_{1}t_{k})+\omega^{2}_{1}}.
 \label{pulsep}
\eeqa
The evolution of the the momentum distribution at different stages of the protocol is shown in Fig. \ref{MainFig3}, for an initially confined TG gas undergoing an implosion engineered by a sudden frequency increase. The oscillatory time-dependence of $b(t)$ shown in Fig. \ref{MainFig2} yields associated oscillations of the momentum distribution, which exhibits DF exactly at the characteristic times $t_{m}=\frac{(2m+1)\pi}{4\omega_{1}}$ with integer $m$. The required pulse strength takes then the maximum value $\left|\dot{b}(t_{m})b(t_{m})\right|=|\omega^{2}_{0}-\omega^{2}_{1}|$. Application of a pulse satisfying the generalized DKC condition (\ref{pulsep}) gives rise to the reversal of DF, rescaling the initial momentum distribution by a factor $1/b(t_k)$; see Fig. \ref{MainFig3}.

In summary, we have established the unitary character of the dynamics of the OBRDM of a TG gas in a driven harmonic trap and point out its far-reaching consequences. The time evolution is exactly isentropic and preserves the occupation numbers of the natural orbitals.
As a result, DF does not affect the distinguishability between the OBRDMs of the dual systems, which is independent of time, e.g., as quantified by the Uhlmann fidelity. 
For arbitrary driving of the trap frequency, the unitary character of the dynamics makes it possible to describe DF as a result of a canonical transformation in phase space, that relates the asymptotic momentum distribution to the initial density profile under rapid acceleration of the width of the atomic cloud. This relation holds for expansions as well as compressions leading to an increase of the interparticle density and is not restricted to the TG gas or systems with hard-core interactions.
Thanks to unitarity, DF can be further tailored and completely reversed by applying a kick with a pulsed external potential, generalizing DKC to interacting systems. This allows to rescale the initial momentum distribution without the requirement of slow driving. Our findings open the way to control nonequilibrium correlations in driven ultracold gases and can be tested in laboratory settings used in recent experiments \cite{Wilson1461}. They should be generalizable to atomic mixtures, systems with $p$-wave interactions, finite coupling strength, fractional exchange statistics, and spinor gases, among other examples.

 {\it Acknowledgements.---}
It is a pleasure to acknowledge discussions with Fernando J. G\'omez-Ruiz, Niklas H\"ornedal, Federico Roccati, Naim Mackel, and Maxim Olchanii.

\let\oldaddcontentsline\addcontentsline   
\renewcommand{\addcontentsline}[3]{}     

\bibliography{DKC_TG_lib} 

\clearpage
\newpage

\onecolumngrid 

\setcounter{equation}{0} 
\setcounter{section}{0}
\setcounter{subsection}{0} 
\global\long\def\theequation{S\arabic{equation}}%
\setcounter{enumiv}{0}

\begin{center}
\textbf{\large{}Supplemental Material for \\``Tailoring Dynamical Fermionization: Delta kick cooling of a Tonks-Girardeau gas''}{\large\par}
\par\end{center}

In this Supplemental Material, we present the time evolution of the OBRDM, the Wigner function for the spin-polarized Fermi gas, the analysis of DF based on saddle point approximation, the analysis of DF based on the Wigner function, and protocols for tailoring DF and Delta kick cooling of a TG gas in a time-dependent harmonic trap.

%

\let\addcontentsline\oldaddcontentsline   

\tableofcontents

\addtocontents{toc}{\protect\thispagestyle{empty}}
\pagenumbering{gobble}

\section{Time-evolution of the OBRDM}

Consider a system in a harmonic trap with frequency $\om_0$ which is modulated as $\om(t)$ for $t>0$.
For scale-invariant dynamics, the wavefunction of an energy eigenstate at $t=0$ evolves for $t>0$ according to \cite{Gritsev10,delcampo11}
\begin{eqnarray}
\Psi(\vec{r}_{1},\dots, \vec{r}_{\N},t)=\frac{1}{b^{\frac{DN}{2}}}\exp\left[i\frac{m\dot{b}}{2\hbar b}\sum_{i=1}^{\N}\vec{r}_{i}\,^{2}-i\int_{0}^{t}\frac{E(0)}{\hbar b^{2}(t')}dt'\right]\Psi \left(\frac{\vec{r}_{1}}{b},\cdots,\frac{\vec{r}_{\N}}{b};0\right),
\end{eqnarray}
where $\N$ is the number of particles, $D$ is the dimension of the trap, and $b(t)$ is the scaling factor determined by the Ermakov equation $\ddot{b}+\om(t)^2b=\om_0^2/b^{3}$. Note that the coordinate-dependent phase factor is common to all quantum states, while $E(0)$ is the eigenvalue of the initial energy eigenstate $\Psi(0)$. In one dimension $D=1$, the time-evolution of a many-particle eigenstate simply reads 
\begin{eqnarray}
\Psi(t)=\frac{1}{b^{\frac{\N}{2}}}\exp\left[i\frac{m\dot{b}}{2\hbar b}\sum_{i=1}^{\N}x^{2}_{i}-i\int_{0}^{t}\frac{E(0)}{\hbar b^{2}(t')}dt'\right]\Psi \left(\frac{x_{1}}{b},\cdots,\frac{x_{\N}}{b},0\right).
\end{eqnarray}
The corresponding one-body reduced density matrix evolves according to
\begin{eqnarray}
\rho_{1}(x,x';t)&=& \N\int d x_{2}\cdots dx_{\N}\Psi(x,x_{2},\cdots,x_{\N},t)\Psi^{*}(x',x_{2},\cdots,x_{\N},t)\nonumber\\
&=&\exp\left[i\frac{m\dot{b}}{2\hbar b}(x^{2}-x'\ ^2)\right]\frac{\N}{b}\int d \frac{x_{2}}{b}\cdots d\frac{x_{\N}}{b}\Psi \left(\frac{x}{b},\frac{x_{2}}{b},\cdots,\frac{x_{\N}}{b},t=0\right)\Psi^{*}\left(\frac{x'}{b},\frac{x_{2}}{b},\cdots,\frac{x_{\N}}{b},t=0\right)\nonumber\\
&=&\exp\left[i\frac{m\dot{b}}{2\hbar b}(x^{2}-x'\ ^2)\right]\frac{1}{b}\rho_{1}\left(\frac{x}{b},\frac{x'}{b},t=0\right).
\end{eqnarray}
 
Using it, the momentum distribution is explicitly given by
\begin{eqnarray}
n(p,t) &=&\la p|\rho_1(t)|p\ra\nonumber\\
&=&\frac{b}{2\pi\hbar} \int dx dx\,'
 \,\exp\left[i\frac{m\dot{b}b}{2\hbar} (x\,^2-x\,'\,^2)\right] e^{-i\frac{pb}{\hbar}(x-x\,')}\rho_1\left(x,x\,',t=0\right).
\end{eqnarray}

Similar considerations apply to higher-order reduced density matrices that can also be efficiently computed using scale-invariance.
For compactness, we introduce the notation $\vec{x}_s=(x_1,\dots,x_s)$, with $s\in\mathbb{N}$. 
The time-evolution of the $s$-{th} reduced density matrix reads
\begin{eqnarray}
\rho_{s}(\vec{x}_s,\vec{y}_s,t)&=&\frac{\N!}{(\N-s)!}\int dx_{s+1}\dots d x_{\N} \ \Psi(\vec{x}_s,x_{s+1},\dots, x_{\N},t)
\Psi(\vec{y}_s,x_{s+1},\dots, x_{\N},t)^{*}.\\
&=&\frac{1}{b^{s}}\exp\left[i\frac{m\dot{b}}{2\hbar b}(\vec{x}_s\,^{2}-\vec{y}_s\,^{2})\right] \rho_{s}\left(\frac{\vec{x}_s}{b},\frac{\vec{y}_s}{b},t=0\right).
\end{eqnarray}

Making use of DKC, upon application of the pulse at $t_k$ with duration $\tau_k$, the dynamical phase is suppressed, and one recovers the same density matrix that would be obtained in an adiabatic protocol
\begin{eqnarray}
\rho_{s}(\vec{x}_s,\vec{y}_s,t_k+\tau_k)=\frac{1}{b^{s}}\rho_{s}\left(\frac{\vec{x}_s}{b},\frac{\vec{y}_s}{b},t=0\right),
\end{eqnarray}
with $b=b(t_k)$.
The joint characteristic function is obtained by the $s$-dimensional Fourier transform, 

\begin{eqnarray}
 n_{s}(\vec{p}_s,t)=\frac{1}{(2\pi\hbar)^{s}}\int d\vec{x}_sd\vec{y}_s e^{-i\sum_{j=1}^{s}\frac{p_{j}(x_{j}-y_{j})}{\hbar}}\rho_{s}(\vec{x}_s,\vec{y}_s,t).\nonumber\\
\end{eqnarray}

After a DKC pulse, the characteristic function reads
\begin{eqnarray}
 n_{s}(\vec{p}_s,t_k+\tau_k)&=&\frac{1}{(2\pi\hbar)^{s}}\int d\vec{x}_s d\vec{y}_s\frac{1}{b^{s}}e^{-i\sum_{j=1}^{s}\frac{p_{j}(x_{j}-y_{j})}{\hbar}}\rho_{s}\left(\frac{\vec{x}_s}{b},\frac{\vec{y}_s}{b},t=0\right),\\
 &=&b^{s} n_{s}(b\vec{p}_s,t=0),
\end{eqnarray}
with $b=b(t_k)$.

From the equation of motions for the OBRDM and the $s$-order reduced density matrix, it is possible to conclude that their evolution is described by the unitary 
\beqa
U^{(s)}(t)=\exp\left[i\frac{m\dot{b}}{2\hbar b}\vec{x}_s^2\right]\exp\left[-i\frac{{\rm ln}b}{2\hbar}(\vec{x}_s\cdot\vec{p}_s+\vec{p}_s\cdot\vec{x}_s)\right].
\eeqa

\section{Representation of the scale-invariant unitary dynamics of the OBRDM by canonical transformation}

As shown in Ref.~\cite{GarciaCalderon80}, any classical canonical transformation on the Wigner function corresponds to a unitary evolution in quantum mechanics, which is represented by a kernel function in the phase space. In particular, it was shown that for a linear canonical transformation,
\begin{eqnarray}
\begin{pmatrix}
\vec{r} \\
\vec{p}
\end{pmatrix}
=
\begin{pmatrix}
\alpha & \beta \\
\gamma & \delta
\end{pmatrix}
\begin{pmatrix}
\vec{r}\,' \\
\vec{p}\,'
\end{pmatrix},
\end{eqnarray}
with $\alpha\delta-\beta\gamma=1$, the kernel function in the quantum mechanical evolution takes a very simple form,
\beqa \label{K_HO}
K(\vec{r},\vec{p}|\vec{r}\,',\vec{p}\,')=\delta^D[\vec{r}\,'-(\alpha \vec{r}+\beta \vec{p})]\delta^D[\vec{p}\,'-(\gamma \vec{r}+\delta \vec{p})]\,,\nonumber\\
\eeqa
where $\delta^D$ the $D$-dimensional delta function. Thus, the evolution of the Wigner function is dictated by \cite{GarciaCalderon80,Schuch_2008}
\beqa
W(\vec{r},\vec{p},t)=\iint d^D\vec{r}\,'d^D\vec{p}\,'K(\vec{r},\vec{p}|\vec{r}\,',\vec{p}\,')W(\vec{r}\,',\vec{p}\,',0)=W(\alpha \vec{r}+\beta \vec{p}, \gamma \vec{r}+\delta \vec{p},0).
\eeqa

Under scale-invariance, the time-dependent Wigner function can be solely written in terms of the initial Wigner function:
\beqa
 W(\vec{r},\vec{p},t)&=& \frac{1}{(\pi\hbar)^D} \int\frac{ d^D \vec{y}}{b^D}\exp\left\{i\frac{m\dot{b}}{2\hbar b}[(\vec{r}-\vec{y})\,^2-(\vec{r}+\vec{y})\,^2]\right\}\rho_1\left(\frac{\vec{r}-\vec{y}}{b},\frac{\vec{r}+\vec{y}}{b},0\right) e^{2i \vec{p}\cdot\vec{y}/\hbar}, \nonumber\\
 &=& \frac{1}{(\pi\hbar)^D} \int\frac{ d^D \vec{y}}{b^D}\rho_1\left(\frac{\vec{r}-\vec{y}}{b},\frac{\vec{r}+\vec{y}}{b}, 0\right) e^{2i \left(b\vec{p}-m\dot{b}\vec{r}\right)\cdot\vec{y}/(b\hbar)},\nonumber\\
 &=&W\left(\frac{\vec{r}}{b},b\vec{p}-m\dot{b}\vec{r},0\right). \label{eq:rescaling}
 \eeqa

 Eqquation~\eqref{eq:rescaling} amounts to the following canonical transformation
 
 \begin{eqnarray}
\begin{pmatrix}
\alpha & \beta \\
\gamma & \delta
\end{pmatrix}
=
\begin{pmatrix}
1/b & 0 \\
-m\dot{b} & b
\end{pmatrix}.
\end{eqnarray}

\section{Wigner function for the spin-polarized Fermi gas}

\begin{figure}[H]
\begin{centering}
\includegraphics[width=0.7\linewidth]{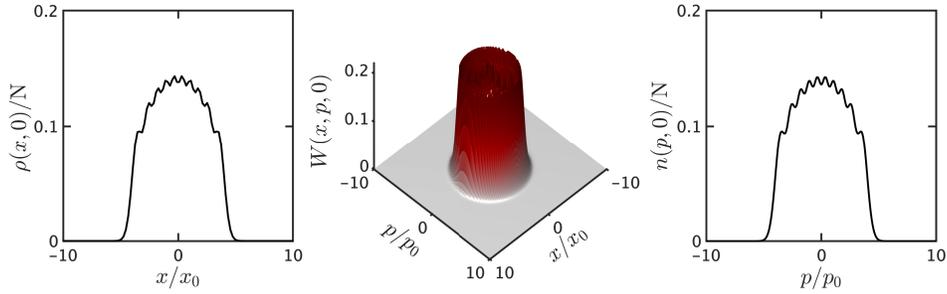}
\caption{Wigner function of ground state of the Fermi gas, in a harmonic trapping of frequency $\omega_{0}$. We note that the ellipso\"idal shape of the Wigner function leads to riddles both in the density profile and in the momentum distribution. \label{Fig:fig4_Wigner_Fermi}}
\end{centering}
\end{figure}

It is of interest to compare the Wigner function of the TG gas in Fig. \ref{Fig: Wigner_density_momentum} to the Wigner function of the dual, ideal Fermi gas plotted in Fig. \ref{Fig:fig4_Wigner_Fermi}.
Given the form of the OBRDM of the ideal Fermi gas $\rho^{\rm F}_{1}(x,x')=\sum_{n=0}^{\N-1}\phi_{n}(x)\phi^{*}_{n}(x')$, it follows that the Wigner function is given 
as
\beqa
W^{\rm F}(x,p)=\sum_{n=0}^{\N-1}W_n(x,p),
\eeqa
in terms of the sum of the Wigner function of single-particle eigenstates of the harmonic oscillator. The latter is well know and takes the form (see e.g. \cite{Shanahan18})
\beqa
W_n(x,p)=\frac{(-1)^n}{\pi\hbar }\exp\left[-\frac{2}{\hbar\om_0}\left(\frac{p^2}{2m}+\frac{1}{2}m\om^2x^2\right)\right]L_n\left[\frac{4}{\hbar\om_0}\left(\frac{p^2}{2m}+\frac{1}{2}m\om^2x^2\right)\right],
\eeqa
where $L_n(x)$ is the Laguerre polynomial.
The Wigner function of an ideal Fermi gas of $\N$ particles in a harmonic trap is shown in Fig. \ref{Fig:fig4_Wigner_Fermi} together with the corresponding marginals: the density profile $\rho^{\rm F}(x)$ and the momentum distribution $n^{\rm F}(p)$.

\section{Analysis of Dynamical Fermionization in time-dependent traps}
 
 As we have seen in the preceding section, the momentum distribution can be found in terms of the Fourier transform of the OBRDM $n(p,t) = \frac{1}{2\pi\hbar}\int dx dx\,'
 \,e^{-ip(x-x\,')/\hbar} \rho_1(x,x\,';t)$. It is useful to make the change of variables $x/b\to x$ in the distribution
\begin{eqnarray}
\label{npt}
n(p,t) &=&\frac{b}{2\pi\hbar } \int dx dx\,' \,e^{i\phi(x,x')}\rho_1\left(x,x\,',t=0\right) \label{eq:momentum_distribution},\\
 \phi(x,x')&=&\frac{m\dot{b}b}{2\hbar} (x\,^2-x\,'\,^2)-\frac{pb}{\hbar}(x-x\,'). \label{eq:phase}
\end{eqnarray}

Previous literature analyzed DF \cite{MinguzziGangardt05} making use of the Stationary Phase Approximation (SPA) \cite{Bleistein86}, even in scenarios where the dynamics is not scale-invariant \cite{Pedri05,delcampo08}. Here, we review this  approach. The phase $\phi(x,x')$ (\ref{eq:phase}) has stationary points $x^{*}=x'^{*}=\frac{ p}{ m \dot{b} }$ so that after Taylor expansion
\begin{eqnarray}
\phi(x,x')&=&\frac{1}{2}\frac{m\dot{b}b}{\hbar}[(x-x^{*})^{2}-(x'-x'^{*})^{2}].\label{eq: phase_expression}
\end{eqnarray}

We introduce the reference length $x_{0}=\sqrt{\frac{\hbar}{m\omega_{0}}}$ to perform the integration in dimensionless units. Note that the OBRDM $\rho_{1}(x,x';t)$ scales as $1/x_{0}$, while the momentum distribution scales as $x_{0}/\hbar$. One can insert the previous expression (\ref{eq: phase_expression}) in the momentum distribution (\ref{eq:momentum_distribution}). Provided that $\frac{1}{2}\frac{m\dot{b}b}{\hbar}\gg 1/x^{2}_{0}$ (i.e., $\dot{b}b\gg 2\omega_{0}$), one can make use of the SPA \cite{Bleistein86} as proposed in \cite{MinguzziGangardt05} to analyze DF. For instance, the free expansion gives $b(t)=\sqrt{1+(\omega_{0}t)^{2}}$, $\dot{b}=\frac{\omega^{2}_{0}t}{\sqrt{1+(\omega^{2}_{0}t)^{2}}}$, implying $\dot{b}b=\omega^{2}_{0}t$ and $t_{DF}\gg\frac{2}{\omega_{0}}$, and leading to the asymptotic momentum distribution
\begin{eqnarray}
n(p,t) &\approx&\frac{b}{2\pi \hbar } \int dx dx\,' \,e^{i\frac{1}{2}\frac{m\dot{b}b}{\hbar}[(x-x^{*})^{2}-(x'-x'^{*})^{2}]}\rho_1\left(x^{*},x'^{*},t=0\right),\nonumber\\
&\approx&\frac{b}{2\pi \hbar } x^{2}_{0}\int \frac{dx}{x_{0}} \frac{dx\,'}{x_{0}} \,e^{i\frac{1}{2}\frac{\dot{b}b}{\omega_{0}}\frac{[(x-x^{*})^{2}-(x'-x'^{*})^{2}]}{x^{2}_{0}}}\rho_1\left(x^{*},x'^{*},t=0\right),\nonumber\\
 &=&\frac{\omega_{0}}{\dot{b}}\frac{x^{2}_{0}}{\hbar}\rho\left(x^{*},t=0\right)=\frac{1}{m\dot{b}}\rho\left(\frac{p}{m\dot{b}},t=0\right).
\end{eqnarray}
This equation can also be demonstrated without relying on scale-invariance when the dynamics is described by the single-particle free-propagator (no interactions) \cite{Pedri05}. 

For the harmonic trap, it is possible to go further and relate the density profile of the TG gas to the momentum distribution of the Fermi gas. This is possible as the OBRDM of the ideal Fermi gas can be directly expressed in terms of the single-particle harmonic-oscillator eigenstates 
\begin{eqnarray}
\rho^{\rm F}_{1}(x,x')=\sum_{n=0}^{\N-1}\phi_{n}(x)\phi^{*}_{n}(x'),
\end{eqnarray}
whence it follows that the momentum distribution reads
 \beqa
 n^{\rm F}(p,0)=\sum_{n=0}^{\N-1}\left|\tilde{\phi}_{n}(p)\right|^2,
 \eeqa
 in terms of the Fourier transform $\tilde{\phi}_{n}(p)$ of the eigenstates $\phi_{n}(x)$ of the Harmonic Oscillator $H=\frac{p^{2}}{2m}+\frac{1}{2}m\omega^{2}_{0}x^{2}$.
 Given that $\tilde{\phi}_{n}(p)=\frac{(-i)^{n}}{\sqrt{m\omega_{0}}}\phi_{n}\left(\frac{1}{m\omega_{0}}p\right)$,
we note that the density profile evaluated at the stationary points reads
\begin{eqnarray}
\rho^{\rm F}(x^{*},t=0)=\sum_{n=0}^{\N-1}\left|\phi_{n}\left(\frac{p}{m\dot{b}},0\right)\right|^{2}=m\omega_{0} \ n^{\rm F}\left(\frac{\omega_{0}}{\dot{b}}p,t=0\right).
\end{eqnarray}

As the wavefunctions of the TG and Fermi gases only differ by the Bose-Fermi mapping, the density profile of the TG and ideal Fermi gas are equal at all times. In particular, they are equal at $t=0$, i.e., $\rho^{\rm TG}(x,t=0)=\rho^{\rm F}(x,t=0)$. Thus, for the harmonic potential, the asymptotic momentum distribution of the TG gas is given in terms of that the Fermi gas, which is the central feature of DF \cite{RigolMuramatsu05,MinguzziGangardt05}
\begin{eqnarray}
n^{\rm TG}(p,t)&\approx&\frac{\omega_{0}}{\dot{b}} n^{\rm F}\left(\frac{\omega_{0}}{\dot{b}}p,t=0\right).
\end{eqnarray}

\section{Dynamical fermionization in phase space: asymptotic marginal distributions of the Wigner function}
The marginals of the Wigner function read 
\begin{eqnarray}
\rho(x,t) & = & \int_{-\infty}^{+\infty} dp\ W\left(\frac{x}{b},bp-m\dot{b}x,t=0\right)=\int_{-\infty}^{+\infty}\frac{dp}{b}\ W\left(\frac{x}{b},p,t=0\right),\\
n(p,t) & = & \int_{-\infty}^{+\infty}dx\ W\left(\frac{x}{b},bp-m\dot{b}x,t=0\right)=\int_{-\infty}^{+\infty}\frac{dk}{m\dot{b}}\ W\left(\frac{p}{m\dot{b}}-\frac{k}{m\dot{b}b},k,t=0\right),\label{eq:n-p}
\end{eqnarray}
where the last identity makes use of the change of variable $k=bp-m\dot{b}x$.
The first equation leads to 
\begin{equation}
\rho(x,t)=\frac{1}{b}\rho\left(\frac{x}{b},t=0\right).
\end{equation}
We further assume $\dot{b}(t)b(t)$ is large and perform a Taylor expansion on the integrand in r.h.s. of Eq.~(\ref{eq:n-p}),
i.e.,
\begin{equation}
W\left(\frac{p}{m\dot{b}}-\frac{k}{m\dot{b}b},k,0\right)=W\left(\frac{p}{m\dot{b}},k,0\right)+\sum_{n=1}^{\infty}\frac{1}{n!}\left(\frac{1}{mb\dot{b}}\right)^{n}W_{1}^{(n)}(x,k,\,0)\big|_{x=p/(m\dot{b})}k^{n},
\end{equation}
we find 
\begin{equation}
n(p,\,t)=\frac{1}{m\dot{b}}\rho\left(\frac{p}{m\omega_{0}},t=0\right)+\frac{1}{m\dot{b}}\sum_{n=1}^{\infty}\frac{1}{n!}\left(\frac{1}{mb\dot{b}}\right)^{n}\int_{-\infty}^{\infty}dkW_{1}^{(n)}(x,k,\,0)\big|_{x=p/(m\dot{b})}k^{n},
\end{equation}
where $W_{1}^{(n)}(x,\,y,t)$ denotes the $n$-th partial derivative
with respect to the first argument. Next, we show that when
\begin{align}
\frac{\partial^{p+q}\rho_{1}(x,\,y)}{\partial^{p}x\partial y^{q}} & \to0,\,\text{as}\,x\to\pm\infty,y\to\pm\infty,\,\forall p,\,q,\label{eq:convergence}
\end{align}
one can find 
\begin{equation}
\left|\int_{-\infty}^{\infty}dkW_{1}^{(n)}(x,k,\,0)k^{n}\right|<\infty,\,\forall n\ge1.\label{eq:high-order}
\end{equation}
Thus at late times, to the first-order approximation, 
\begin{equation}
n(p,t)\approx\frac{1}{m\dot{b}}\rho\left(\frac{p}{m\omega_{0}},t=0\right).
\end{equation}

To show Eq.~(\ref{eq:convergence}) implies Eq.~(\ref{eq:high-order}),
we first resort to the definition of the Wigner transformation and
find 
\begin{align}
\int_{-\infty}^{\infty}dkW_{1}^{(n)}(x,k,\,0)k^{n} & =\frac{1}{\pi\hbar}\int_{-\infty}^{\infty}dy\int_{-\infty}^{\infty}dk\frac{\partial^{n}\rho_{1}(x-y,\,x+y)}{\partial x^{n}}e^{2\text{i}ky/\hbar}k^{n}\nonumber \\
 & =\frac{1}{\pi\hbar}\left(\frac{\hbar}{2i}\right)^{n}\int_{-\infty}^{\infty}dy\int_{-\infty}^{\infty}dk\frac{\partial^{n}\rho_{1}(x-y,\,x+y)}{\partial x^{n}}\frac{\partial^{n}e^{2\text{i}ky/\hbar}}{\partial y^{n}}.
\end{align}
Integration by parts yields
\begin{equation}
\int_{-\infty}^{\infty}dkW_{1}^{(n)}(x,k,\,0)k^{n}=\frac{1}{\pi\hbar}\left(\frac{\hbar}{2i}\right)^{n}\int_{-\infty}^{\infty}dy\int_{-\infty}^{\infty}dk\frac{\partial}{\partial y}\frac{\partial^{n}\rho_{1}(x-y,\,x+y)}{\partial x^{n}}\frac{\partial^{n-1}e^{2\text{i}ky/\hbar}}{\partial y^{n-1}},
\end{equation}
where the boundary term vanishes since
\begin{equation}
\lim_{y\to\pm\infty}\frac{\partial^{n}\rho_{1}(x-y,\,x+y)}{\partial x^{n}}=0,
\end{equation}
according to Eq.~(\ref{eq:convergence}). Iterating this process
for $n-1$ times, we find 
\begin{align}
\int_{-\infty}^{\infty}dkW_{1}^{(n)}(x,k,\,0)k^{n} & =\frac{1}{\pi\hbar}\left(\frac{\hbar}{2i}\right)^{n}\int_{-\infty}^{\infty}dy\int_{-\infty}^{\infty}dk\frac{\partial^{n}\rho_{1}(x-y,\,x+y)}{\partial x^{n}\partial y^{n}}e^{2\text{i}ky/\hbar}\nonumber \\
 & =\left(\frac{\hbar}{2i}\right)^{n}\int_{-\infty}^{\infty}dy\frac{\partial^{n}\rho_{1}(x-y,\,x+y)}{\partial x^{n}\partial y^{n}}\delta(y)\nonumber \\
 & =\left(\frac{\hbar}{2i}\right)^{n}\frac{\partial^{n}\rho_{1}(x-y,\,x+y)}{\partial x^{n}\partial y^{n}}\bigg|_{y=0},
\end{align}
which immediately implies Eq.~(\ref{eq:high-order}).

Finally, we note for an initial eigenstate of the TG gas and the spin-polarized
fermi gas, according to the definition of the OBRDM, 
\begin{equation}
\rho_{1}(x,y,0)=\N\int dx_{2}\dots dx_{\N}\ \Psi(x,x_{2},\dots,x_{\N},0)\Psi^{*}(y,x_{2},\dots,x_{\N},0).
\end{equation}
In the case of harmonic confinement, Eq. (\ref{eq:convergence}) is guaranteed by the
property of the many-body wave function, 
\begin{equation}
\partial^{n}\Psi(x,x_{2},\dots,x_{\N},0)/\partial x^{n}\to0,\,\text{as},\,x\to\pm\infty,\,\forall n.
\end{equation}

\section{Tailoring Dynamical Fermionization and Delta kick cooling of a Tonks-Girardeau gas in a time-dependent harmonic trap}
\subsection{Ground-state OBRDM of a TG gas in a harmonic trap}
To compute the evolution of the momentum distribution (\ref{eq:momentum_distribution}) it is necessary to determine the initial density matrix explicitly. A general expression for a pure state was derived by Pezer and Buljan making use of the Bose-Fermi mapping \cite{Pezer07}. Under harmonic confinement, further progress is possible, recognizing that the Slater determinant is that of a Vandermonde matrix. Closed formulas of the integrals can be found in terms of gamma and hypergeometric functions \cite{Lapeyre2002, Forrester2003}. 
This expression is useful for the numerical evaluation of the momentum distribution and we use it in our work.
The ground state wavefunction of a harmonically trapped TG gas can be simply written as
\beqa
\Psi_{\rm TG}(x_1,\dots,x_N)=\frac{1}{\sqrt{\N!}}\left|\det_{n=0,k=1}^{\N-1,\N}[\phi_n(x_k)]\right|,
\eeqa
in terms of the well-known eigenfunctions of the single-particle harmonic oscillator $\phi_n(x_k)$.
Recognizing that Slater determinant involved is that of a Vandermonde matrix, one finds that the ground-state wavefunction admits the Jastrow form \cite{Girardeau01} taking $m=\omega_{0}=1$
\beqa
\Psi_{\rm TG}(x_1,\dots,x_N)=C_{\N}\exp\left(-\sum_j\frac{x_j^2}{2x_0^2}\right)\prod_{j<k}|x_j-x_k|,
\eeqa
with normalization constant
\beqa
C_{\N}=\frac{2^{\frac{\N(\N-1)}{4}}}{x_0^{\N/2}\left(N!\pi^{\frac{\N}{2}}\prod_{n=0}^{\N-1}n!\right)^{1/2}}.
\eeqa
As a result, in units of $x_0$, the one-body density matrix reads \cite{Forrester2003}
\begin{eqnarray}
\rho_{1}(x,y)&=&\N\int_{-\infty}^{+\infty}dx_{1}\cdots\int_{-\infty}^{+\infty}dx_{\N-1}\Psi_{\N}(x_{1},\cdots,x_{\N-1},x) \Psi_{\N}(x_{1},\cdots,x_{\N-1},y)^{*}\\
&=&\frac{\N \ 2^{\N-1}}{(\N-1)!\sqrt{\pi} }e^{-(x^{2}/2)-(y^{2}/2)}\int_{-\infty}^{+\infty}dx_{1}\cdots\int_{-\infty}^{+\infty}dx_{\N-1}\prod_{l=1}^{\N-1}e^{-x^{2}_{l}}|x-x_{l}|\nonumber\\
&& \times|y-x_{l}|({\rm det}[\frac{1}{ \sqrt{\sqrt{\pi} 2^{j-1}}\ (j-1)! }x^{j-1}_{k}]_{j,k=1,\dots,\N-1})^{2}\\
&=&\frac{\N\ 2^{\N-1}}{(\N-1)!\sqrt{\pi} }e^{-(x^{2}/2)-(y^{2}/2)}\int_{-\infty}^{+\infty}dx_{1}\cdots\int_{-\infty}^{+\infty}dx_{\N-1}\prod_{l=1}^{\N-1}e^{-x^{2}_{l}}|x-x_{l}|\nonumber\\
& &\times|y-x_{l}|\frac{1}{\N!}\sum_{P\in S_{\N-1}}\epsilon(P)\sum_{Q\in S_{\N-1}}\epsilon(Q)x^{P(l)-1}_{l}x^{Q(l)-1}_{l}\\
&=&\frac{\N\ 2^{\N-1}}{(\N-1)!\sqrt{\pi} }e^{-(x^{2}/2)-(y^{2}/2)}\frac{1}{\N!}\sum_{P\in S_{\N-1}}\epsilon(P)\sum_{Q\in S_{\N-1}}\epsilon(Q)\prod_{l=1}^{\N-1}\int_{-\infty}^{+\infty}dx_{l}e^{-x^{2}_{l}}|x-x_{l}|\nonumber\\
& & \times|y-x_{l}|\ \frac{1}{ \sqrt{\sqrt{\pi} 2^{P(l)-1}}\ (P(l)-1)! } \frac{1}{ \sqrt{\sqrt{\pi} 2^{Q(l)-1}}\ (Q(l)-1)! } \ x^{P(l)-1}_{l}\ x^{Q(l)-1}_{l}\\
&=&\frac{2^{\N-1}}{\Gamma(\N) \sqrt{\pi} } e^{-(x^{2}/2)-(y^{2}/2)}{\rm det}\left[\frac{2^{(j+k)/2}}{2\sqrt{\pi}\sqrt{\Gamma(j)\Gamma(k)}}b_{j,k}(x,y)\right]_{j,k=1,\cdots,\N-1}\, ,
\end{eqnarray}
with 
\begin{eqnarray}
b_{j,k}(x,y)=\int_{-\infty}^{+\infty}dt \ e^{-t^{2}}|x-t| \ |y-t| \ t^{j+k-2}. \label{eq:bjk}
\end{eqnarray}
Note that this provides the one-body reduced density matrix at $t=0$. Its subsequent time evolution for $t>0$ can be described by making use of this expression and scale invariance. 
 The corresponding momentum distribution can be readily obtained via Fourier transformation.
 
\subsection{Delta kick cooling of an expanding TG gas}
 \begin{figure}[H]
 \begin{centering}
\includegraphics[width=0.4\linewidth]{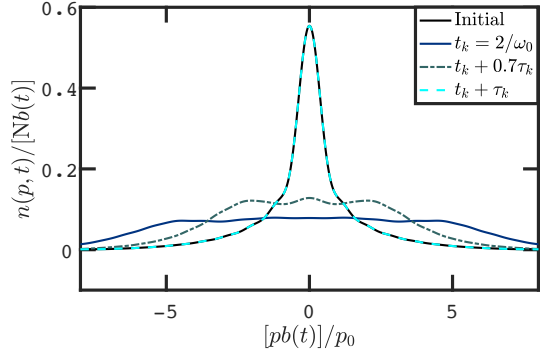}
\caption{DKC of an expanding TG gas. (a) Momentum distribution $n(p)/[\N b(t)]$ where $\N$ is the number of particles and $b(t)$ the scaling factor solution to the Ermakov equation.
The initially peaked momentum distribution broadens with the expansion as a result of DF, approaching that of an ideal Fermi gas, at $2/\omega_{0}$. A general kick, such as that with kick strength $0.7\tau_{k}$, further distorts the momentum distribution. Only when the kick duration is chosen according to the DKC relation, DF is completely reversed, and the momentum distribution of the final state matches that of an adiabatic expansion and is given by that at $t=0$, up to scaling ($\N=7$). 
\label{SMFigTGexp}}
\end{centering}
\end{figure}
We consider a cloud of atoms that is initially trapped in a harmonic potential of frequency $\omega_{0}$. At time $t=0$ the trap is released so that $\omega(t>0)=0$ and the solution of the Ermakov equation is given by 
$b(t)=\sqrt{1+(\omega_{0}t)^{2}}$.
 Fig. \ref{SMFigTGexp} depicts the momentum distribution for the TG gas trapped in a harmonic potential at initial time $t=0$, its evolution after a free expansion governed by DF, and after a kick of duration $\tau_{k}$, reversing DF. One can notice that the kick allows recovering a distribution of bosonic shape. However, it is interesting to notice that the distribution is rescaled compared to the initial momentum distribution, following equation (\ref{eq:momentum_distribution}), the expansion followed by the kick acting as a microscope. This rescaling matches with the expression found in \cite{MinguzziGangardt05} predicting the evolution of the distribution for large $p$ as $n(p,t)\approx p^{-4}b^{-3}$
 that matches with the prediction of a decaying tail upon completion of the DKC protocol $n_{\rm kick}(p,t)=b\times n(pb,0)\approx b \times (pb)^{-4}$.

\subsection{Delta kick cooling of an imploding TG gas}
Intuitively the DF occurs for a trap expansion when the particles get more distant from each other and do not interact. We next show that it is also possible to observe the DF in the less trivial case of a compression process with $b(t_{F})<b(t=0)=1$ if the process is short enough so that the product $\dot{b}b$ is large.
The implosion can be engineered by a modulation of the trapping potential. 
We proceed by reverse engineering. We first set a desired time evolution of the scaling factor. For the sake of illustration, we consider
\beqa
\label{bimplo}
b(t)=1+(b_{F}-1)\left(\frac{t}{t_{F}}\right)^{3}.
\eeqa

Making use of the Ermakov equation, the time-dependent frequency $\om^2(t)$ that leads to thus time-dependence of $b(t)$ can be determined, as shown in Fig. \ref{Fig2}. 
Large values of $\dot{b}b$ induce a high-frequency phase modulation in the coordinate representation. 
\begin{figure}[t]
\begin{centering}
\includegraphics[width=0.7\linewidth]{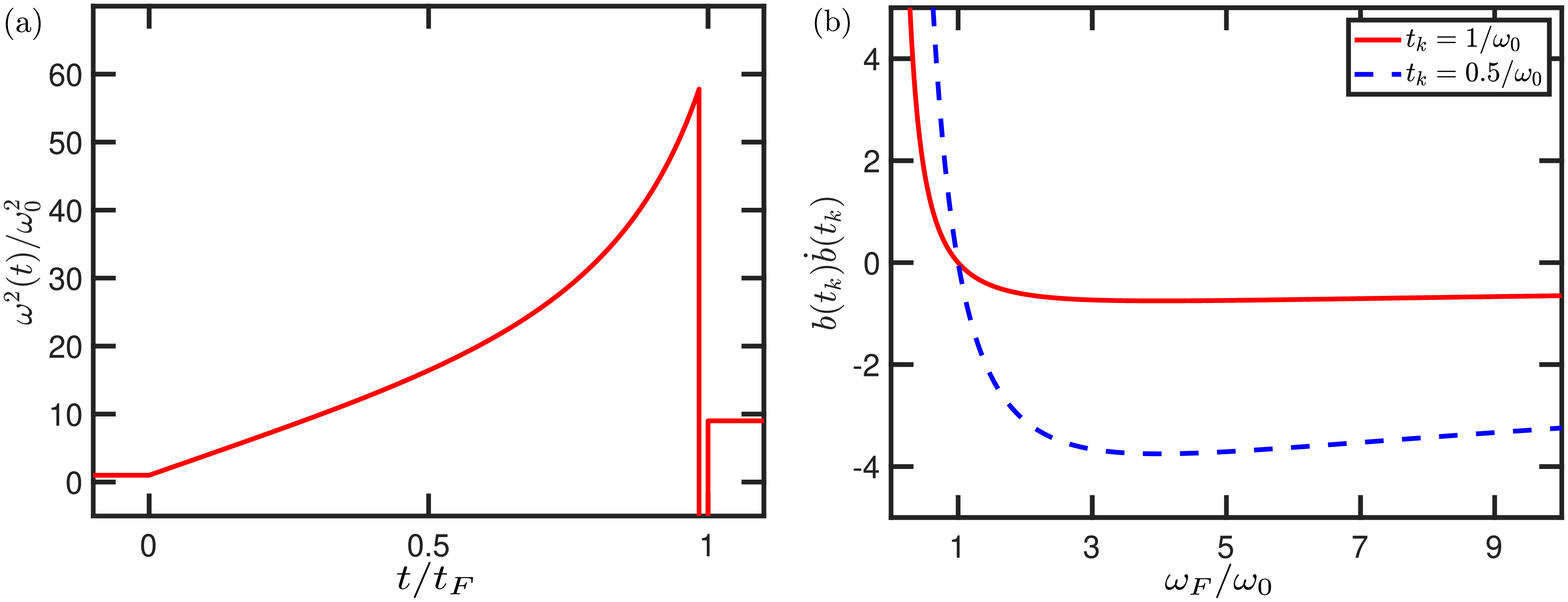}
\caption{ (a) The frequency modulation leading to the implosion protocol (\ref{bimplo}), involving a smooth quench at $t=0$ and a subsequently smooth modulation. DF is reversed in an implosion making use of a pulsed inverted oscillator. We chose the frequency of the kick $\omega^{2}_{k}=1000\omega^{2}_{0}$, the compression time protocol as $t_{k}=0.3/\omega_{0}$ and the final frequency determined by $b_{F}=\sqrt{\omega_{0}/\omega_{F}}=\sqrt{1/3}$. (b) Strength of the phase modulation shown as a function of the frequency ratio. \label{Fig2}}
\end{centering}
\end{figure}
One may wonder whether it is possible to cancel DF by DKC. 
However, now the derivative of the scaling parameter $\dot{b}$ is negative during the process as shown in Fig. \ref{Fig2}. Also, to cancel the DF it is necessary to apply a kick with an inverted potential. One can now consider the ``kicked'' Hamiltonian where the kick-pulse is repulsive $\om_k=i\om_I$, i.e.,
$H_k(t)=H(t)-\delta(t-t_k)\frac{1}{2}m\om_I^2\sum_{i=1}^{\N} x_i\,^{2}$.
The corresponding time-evolution operator reads
as
$
U_{ \delta}(t_{k},0)=\exp\left(+i\frac{\tau_k m\om_I^2}{2\hbar}\sum_{i=1}^{\N} x_i\,^{2}\right)U(t_k,0)$,
where $U(t_{k}, 0)$ is the propagator associated with $H(t)$, and $\tau_k$ is a small time scale during which the kick is applied. As a result, the relation for the pulse parameters to reverse DF is still given by Eq. (\ref{DKCcond}). For the prescribed trajectory (\ref{bimplo}), applying the kick at the time $t_k$ and duration $\tau_k$, the required pulse parameters to reverse DF are set by
\beqa
 \tau_{k}\omega^{2}_{k}=3\frac{(b_{F}-1)}{t_{k}b_{F}}.
\eeqa

\begin{figure}
\includegraphics[width=0.4\linewidth]{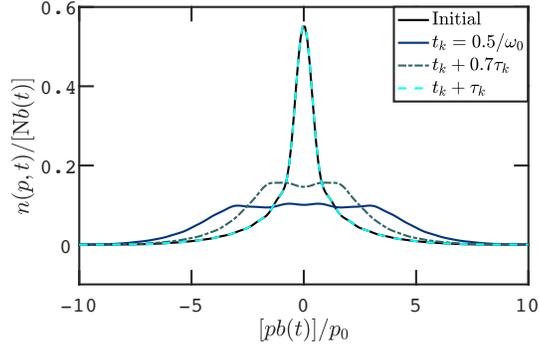}
\caption{DKC of an imploding TG gas. (a) Momentum distribution $n(p)$ of a trapped gas, after the implosion at $t_{k}=0.5/\omega_{0}$ and after the delta-kick for $0.7 \tau_{k}$ (green dashed-dotted line), after the delta-kick $\tau_{k}$ (turquoise blue dashed line); $\N=7$. In dashed green the kick strength is chosen as $\tau_{k}\omega^{2}_{k}$ in order to cancel DF, with $\omega^{2}_{k}=1000 \omega^{2}_{0}$. In choosing the strength of the kick differently one can modulate DF. The implosion protocol is described by Eq. (\ref{bimplo}) with $b_{F}=\sqrt{1/5}$. 
 \label{Fig2}}
\end{figure}

\end{document}